\documentclass[prl,aps,twocolumn,superscriptaddress,floatfix,notitlepage,nofootinbib,amssymb,amsmath]{revtex4-1}
\usepackage{array,multirow,graphicx}
\usepackage{mathtools}
\usepackage{epsfig}

\usepackage{xcolor}
\usepackage{ulem}
\definecolor{purple}{rgb}{0.8,0,0.6}

\newcommand{\ed}[1]{#1}
\newcommand{\beqn}{\begin{eqnarray}}
\newcommand{\eeqn}{\end{eqnarray}}
\newcommand{\eq}[1]{(\ref{#1})}

\newcommand{\cS}{{\cal S}}
\newcommand{\cL}{{\cal L}}

\newcommand{\Cas}{{\mathrm{Cas}}}

\newcommand{\bs}{\boldsymbol}

\newcommand{\avr}[1]{{\left\langle #1 \right\rangle}}

\begin{document}

\title{Casimir effect with machine learning}

\author{M. N. Chernodub}
\affiliation{Institut Denis Poisson UMR 7013, Universit\'e de Tours, 37200 France}
\affiliation{Pacific Quantum Center, Far Eastern Federal University, Sukhanova 8, Vladivostok, 690950, Russia}
\author{Harold Erbin}
\affiliation{Dipartimento di Fisica, Universit\`a di Torino, INFN Sezione di Torino and Arnold-Regge Center, Via Pietro Giuria 1, I-10125 Torino, Italy}
\author{I. V. Grishmanovskii}
\affiliation{Pacific Quantum Center, Far Eastern Federal University, Sukhanova 8, Vladivostok, 690950, Russia}
\author{V. A. Goy}
\affiliation{Pacific Quantum Center, Far Eastern Federal University, Sukhanova 8, Vladivostok, 690950, Russia}
\author{A. V. Molochkov}
\affiliation{Pacific Quantum Center, Far Eastern Federal University, Sukhanova 8, Vladivostok, 690950, Russia}

\begin{abstract}
Vacuum fluctuations of quantum fields between physical objects depend on the shapes, positions, and internal composition of the latter. For objects of arbitrary shapes, even made from idealized materials, the calculation of the associated zero-point (Casimir) energy is an analytically intractable challenge. We propose a new numerical approach to this problem based on machine-learning techniques and illustrate the effectiveness of the method in a (2+1) dimensional scalar field theory. The Casimir energy is first calculated numerically using a Monte-Carlo algorithm for a set of the Dirichlet boundaries of various shapes. Then, a neural network is trained to compute this energy given the Dirichlet domain, treating the latter as black-and-white pixelated images. We show that after the learning phase, the neural network is able to quickly predict the Casimir energy for new boundaries of general shapes with reasonable accuracy. 
\end{abstract}

\date{\today}

\maketitle

The presence of physical bodies in a quantum vacuum affects the spectrum of zero-point fluctuations of quantum fields and leads to the appearance of the forces acting on the bodies. This phenomenon, known as ``the Casimir effect'', was first predicted in 1948 by Hendrik Casimir, who has shown that two strictly parallel neutral metallic plates should attract each other due to quantum fluctuations of the electromagnetic field~\cite{ref:Casimir}. The Casimir phenomenon generalizes the van der Waals interactions between neutral bodies~\cite{Casmir:1947hx} and plays an essential role in microelectromechanical and microfluidic systems at submillimeter scales, where the zero-point forces induced by the quantum fluctuations of electromagnetic fields become significant~\cite{ref:Milton,ref:Bogdag,ref:Rodriguez}.

Geometrical shapes and material composition of physical bodies affect the Casimir forces significantly. Accurate analytical calculations work for a limited set of relatively simple geometries, where the spectrum of vacuum fluctuations is precisely known. The approximate proximity-force calculations~\cite{ref:proximity} may access perturbations around known configurations including near-planar geometries. The Casimir forces and associated energies are also computed with the help of various numerical and semi-analytical techniques~\cite{Johnson:2010ug}, which include methods of the scattering theory~\cite{ref:scattering:1,ref:scattering:2}, factorization~\cite{ref:factorization} and discretization~\cite{ref:discrete} approximations, \ed{worldline approaches}~\cite{Gies:2006cq}, and lattice field theories~\cite{ref:Oleg,ref:paper:12}. 

Experimentally, the Casimir force has been successfully measured with ever increasing precision in various geometries~\cite{ref:experiment:1,ref:experiment:2,ref:experiment:3,ref:experiment:4,ref:experiment:5}.

In our paper, we propose to tackle the complicate problem of calculation of the Casimir energy in general geometries using the Machine learning (ML) approach. The ML technique is a collection of powerful programming tools that allow the computer to find how to perform a task without being explicitly programmed (see~\cite{ref:review:1,ref:review:2} for physicist reviews). In recent years, the ML has revolutionized many fields of engineering and sciences thanks to several breakthroughs, in particular in the design of neural networks. While the neural networks may be slow in training, their predictions are usually coming very fast. Neural networks find increasingly important implementation in the successful investigation of many complex physical systems that involve a large number of degrees of freedom. The non-exhaustive list of the relevant examples includes open quantum systems with high-dimensional Hilbert spaces~\cite{ref:open:1,ref:open:2,ref:open:3}, topological phases in the context of topological band insulators~\cite{ref:top:invariants:1} and field theories~\cite{ref:top:invariants:2, ref:qft:1}, as well as phase structure of many-body, strongly-correlated and field systems in general~\cite{ref:phases:1,ref:phases:2,ref:phases:3,ref:phases:4, ref:phases:5,ref:phases:6,ref:phases:7,ref:phases:8,ref:phases:9,ref:phases:10, ref:phases:11, ref:phases:12}.

We will use the so-called supervised learning procedure, which -- in a very general sense -- consists of establishing a map from some inputs to some outputs by training a neural network on a broad set of known examples. In our case, the inputs are the boundaries imposed on quantum fields, and the outputs are the Casimir energy of the quantum fields in the space with these boundaries. Technically, we illustrate the effectiveness of the method in a field theory with the simplest, Dirichlet boundary conditions. We describe the Dirichlet boundary geometries as black and white images, and then we employ standard neural-network techniques used in the computer vision~\cite{Chollet:2017:DLP} in order to ``recognize'' the correct Casimir energy for a particular shape of boundaries. 

One of the simplest and, at the same time, practically relevant realizations of the Casimir effect appears in photodynamics, the theory of a single Abelian gauge field $a_\mu$ described by the Lagrangian:
\beqn
\cL_{U(1)} = - \frac{1}{4} f_{\mu\nu} f^{\mu\nu}, \qquad f_{\mu\nu} = \partial_\mu a_\nu - \partial_\nu a_\mu\,.
\label{eq:L}
\eeqn
The photodynamics respects $U(1)$ gauge symmetry, $a_\mu(x) \to a_\mu(x) + \partial_\mu \omega(x)$, and possesses, in $(d+1)$ space-time dimensions, $d-1$ physical degrees of freedom. 

The Lagrangian~\eq{eq:L} describes a very simple system of a single non-interacting vector field. However, the nontriviality of the Casimir effect comes from the boundary conditions, namely, from a nontrivial dependence of the (regularized) energy of the quantum fluctuations on the shape of the boundaries of a physical object immersed into the vacuum of the photons~\eq{eq:L}. It is the shape-dependence which makes the problem difficult. 

We address the problem of the shape-dependence of the Casimir energy using a machine-learning approach. Since we are concerned with the proof-of-principle result, it is sufficient to work in two spatial dimensions which is the lowest dimension appropriate for our purposes. 

In two spatial dimensions, the photon has one physical degree of freedom. Restricting ourselves to an idealized case, one could consider an object made of a perfect electric conductor. At its boundary, the tangential component of the electric field is vanishing:
\beqn
\epsilon^{\mu\alpha\beta} n_\mu(x) f_{\alpha\beta} (x) {\biggl|}_{x \in \cS} = 0\,,
\label{eq:U1:BC}
\eeqn
where $n_\mu(x)$ is a vector normal to the boundary at the point~$x$ of the (piecewise) one-dimensional boundary $\cS$, and $f_{\alpha\beta}$ is given in Eq.~\eq{eq:L}. For two parallel static straight wires separated by the distance $R$, the vacuum fluctuations lead to an attractive potential~\cite{Ambjorn:1981xw}:
\beqn
V_\Cas(R) = - \frac{\zeta(3)}{16 \pi R^2},
\label{eq:R}
\eeqn
where $\zeta(x)$ is the zeta function, $\zeta(3) \simeq 1.20206$. 

The problem may be simplified even further by considering the model of a free real-valued field $\phi $:
\beqn
\cL = \frac{1}{2} \partial_\mu \phi  \partial^\mu \phi.
\label{eq:L:phi}
\eeqn
Similarly to photodynamics in two spatial dimensions, this model has one degree of freedom. Instead of \eq{eq:U1:BC}, the boundary may be set by a simpler, Dirichlet condition:
\beqn
\phi(x) {\biggl|}_{x \in \cS} = 0.
\label{eq:BC}
\eeqn
\ed{We naturally recover the result~\eq{eq:R} for the Casimir energy for two parallel wires with the boundaries~\eq{eq:BC}. For a general shape of the boundary, the Casimir problem cannot be treated analytically.}

\ed{In two spatial dimensions, any configuration of boundaries may be treated as a pixelated black-and-white image, in which white pixels correspond to the free unoccupied space while black pixels encode the positions of the Dirichlet boundaries.} The scalar field freely fluctuates in the white spaces and vanishes at the black pixels at which the Dirichlet condition~\eq{eq:BC} is imposed. \ed{
We consider pixel-thin 
static 
boundaries for which the particle creation is absent and the Casimir energy is a time-independent quantity.
}

\ed{Given the simplicity of the model~\eq{eq:L:phi}, and the complexity of the Casimir problem, our approach exposes the advantages of the sophisticated method of the machine-learning approach in the best way. In more realistic cases, the Casimir problem becomes evidently more complicated. For example, the interaction of non-parallel surface segments in $(3+1)d$ photodynamics is affected by a complex mixing of different photon modes that satisfy distinct boundary conditions at the surfaces.} 

We numerically calculate the Casimir energy of the wires of various shapes using the first-principles methods of lattice gauge theory developed earlier in Refs.~\cite{ref:paper:12,ref:paper:3,ref:paper:YM}. The discretized version of the scalar gauge theory~\eq{eq:L:phi} is given by the partition function
\beqn
Z = \prod_x \int\nolimits_{-\infty}^{+\infty} d \phi_x \, e^{- S[\phi]}.
\label{eq:Z}
\eeqn
The integration goes over the field $\phi_x \in {\mathbb R}$ defined on the sites  ${\bs x} \equiv (x_1,x_2,x_3)$ of the Euclidean cubic lattice $L_s^3$ with periodic boundary conditions in all three directions ($0 \leqslant x_\mu \leqslant L_s-1$). The spatial coordinates ($x_1$ and $x_2$) as well as the Wick-rotated imaginary-time ($x_3 \equiv i t$) are of the same length, corresponding to a zero temperature.

In the lattice action of the $d+1$ dimensional model~\eq{eq:L:phi},
\beqn
S[\phi] = \frac{1}{2} \sum_{\bs x} \sum_{\mu = 1}^{d+1} \left( \phi_{{\bs x} + \hat\mu} - \phi_{{\bs x}}\right)^2,
\label{eq:S}
\eeqn
the derivatives are represented by the finite differences. Here $\hat\mu$ is a unit lattice step in the $\mu$th direction. A lattice generalization of the Dirichlet boundary condition~\eq{eq:BC} is straightforward.

The energy of vacuum fluctuations of the scalar field is related to a local expectation value of its energy density,
\beqn
T^{00}_M = \frac{1}{2} \left[ \left( \frac{\partial \phi}{\partial {\bs x}} \right)^2  + \left( \frac{\partial \phi}{\partial t} \right)^2\right],
\label{eq:T00:Minkowski}
\eeqn
with the following Wick-rotated discretized counterpart:
\beqn
T^{00}_E = \frac{1}{4} \sum_{\mu=1}^3 \eta_\mu \left[ \left( \phi_{{\bs x} +\hat\mu} - \phi_{{\bs x}}\right)^2 + \left( \phi_{{\bs x}} - \phi_{{\bs x} - \hat\mu}\right)^2 \right],
\qquad 
\eeqn
where $\eta_1 = \eta_2 = + 1$ and $\eta_3 = - 1$.

The regularized energy density is formally given by 
\beqn
{\cal E}_\cS(x) = \avr{T^{00}_E(x)}_{\cS} - \avr{T^{00}_E(x)}_{0} 
\label{eq:E:norm}
\eeqn
where the subscripts ``0''  and ``$\cS$''  indicate that the expectation value is taken, respectively, in the absence and in the presence of the boundaries $\cS$ of the physical objects. The ultraviolet divergences cancel in Eq.~\eq{eq:E:norm} so that ${\cal E}_\cS(x)$ provides us with a local finite quantity, the Casimir energy density, which is equal to a change in the energy density of the vacuum fluctuations due to the presence of the boundaries~$\cS$. The total Casimir energy is given by an integral of the energy density~\eq{eq:E:norm} over the whole spatial volume.

\ed{
We considered two types of boundary conditions: closed non-self-intersecting lines and two quasi-parallel nonintersecting lines that represent, respectively, deformed circles and corrugated wires. A related problem, the Casimir energy of the scalar field in cavities of rectangular shapes, was discussed in details in Ref.~\cite{Ambjorn:1981xw}.
We would like to notice that the calculation of the Casimir energy has certain important subtleties related to surface counter-terms that regularize the infinite divergences of idealized boundary conditions for closed boundaries. This uncertainty does not affect the force between rigid bodies as well as the Casimir energy of the real materials. We refer the reader to Refs.~\cite{Graham:2002xq,Graham:2002fw,Graham:2003ib} for further discussions on the important interpretation of the Casimir energy in idealized as contrasted to real materials. In line of the main aim of our paper we treat both closed and quasi-parallel lines at the same footing. For uniformity of our presentations, we call both these quantities the ``Casimir energy'' having in mind the subtlety in their physical meaning mentioned above.}

To diversify our efforts, we also considered a technically similar and equally difficult problem of calculation of the mean total action $\avr{S}$ in the two-dimensional Euclidean model given, formally, by Eq.~\eq{eq:S} with $d=1$. Although this quantity has no straightforward physical interpretation, its calculation is as difficult as the calculation of the Casimir energy in 2+1 dimensional model. Below we will apply the very same ML technique to this two-dimensional model. To shorten notations, we will call these models below as $3d$ and $2d$, respectively.

In our numerical simulations we use the methods successfully adopted for studies of the Casimir forces in Abelian gauge theories in Refs.~\cite{ref:paper:12,ref:paper:3,ref:paper:YM}. To calculate the Casimir energy~\eq{eq:E:norm} in the (2+1) dimensional model, we discretize each geometry of the boundaries at $255^3$ lattice and then generate $2\times 10^5$ scalar field configurations using a Hybrid Monte Carlo algorithm which combines standard Monte-Carlo methods~\cite{ref:20} with the molecular dynamics approach. The latter incorporates a second-order minimum norm integrator~\cite{ref:21}. We skip first $10^5$ configurations to assure their thermalisation and subsequent  $10^5$ configurations for the statistical analysis. We employ the same techniques to calculate the mean action in the two-dimensional model, using $10^6$ configurations at $256^2$ lattices. \ed{The method also allows us to calculate the energy density and pressure around the boundaries. In this work, we concentrate only on the single scalar quantity, the Casimir energy~\eq{eq:E:norm}.}

\begin{figure}[ht]
	\centering
	\includegraphics[scale=0.5]{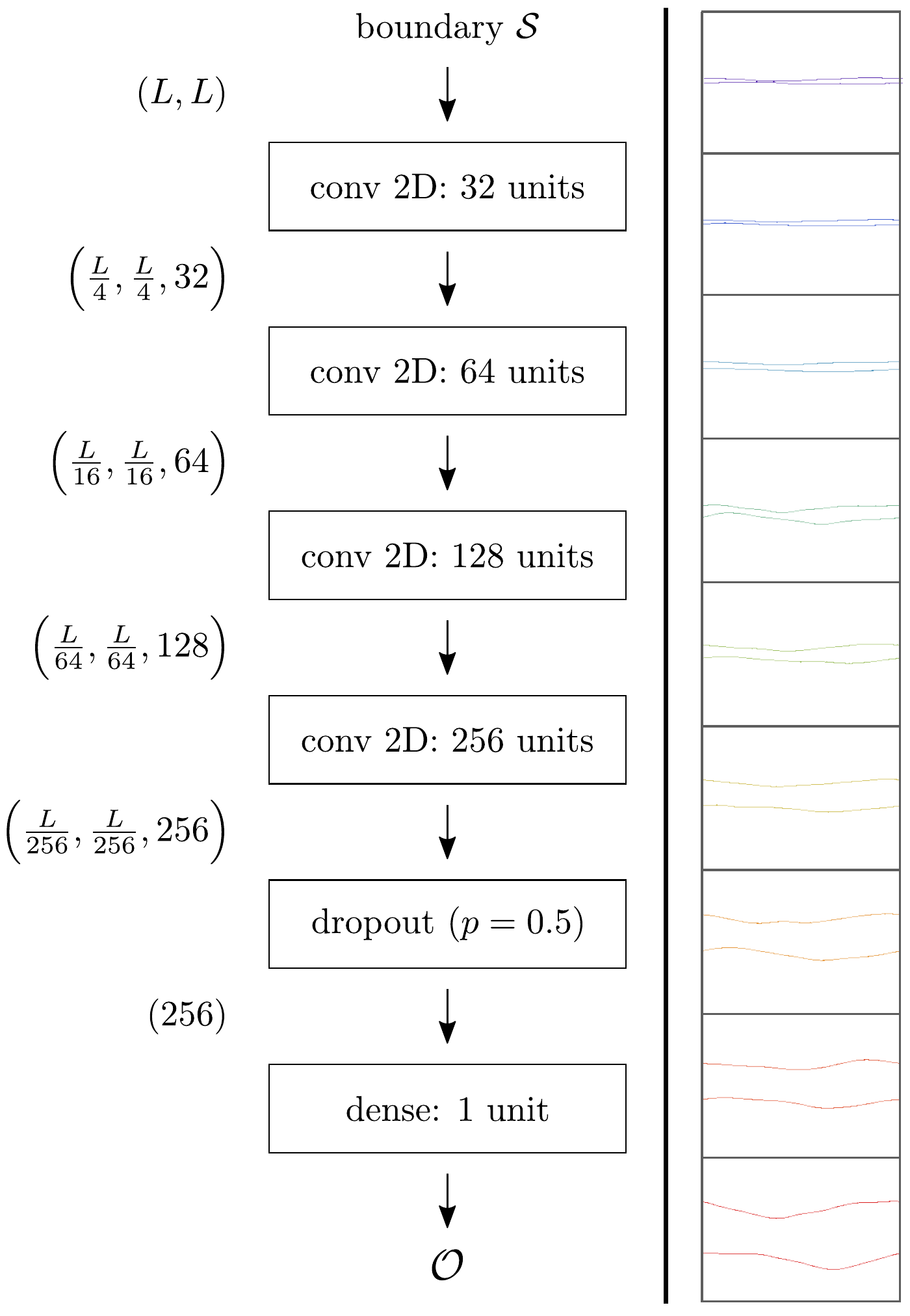}
	\caption{%
		(left) The neural network used to predict the Casimir energy ${\cal E}_{{\cal S}}$ for the static boundary $\cal S$ placed at the spatial $L \times L$ lattice cross-section of the $3d$ model.
		The size of the inputs of each layer is indicated on the left.  The same network is employed for the mean action in the $2d$ model. 
		(right) The examples of the quasi-parallel, corrugated lines used for training and prediction.}
	\label{fig:nn-model}
\end{figure}

The neural network is standard in the context of image processing (Figure~\ref{fig:nn-model}). It is made of four $2d$ convolutional layers with a kernel of size $(3, 3)$ and with $32$, $64$, $128$ and $256$ filters respectively.
Each layer is followed by 
\begin{itemize}
    \item[1)] a batch normalization layer with momentum $0.9$,
    \item[2)] a leaky ReLU activation layer with $\alpha = 0.3$, and 
    \item[3)] a max pooling layer with size $(4, 4)$.
\end{itemize}
The last pooling operation is global in order to collapse the spatial dimensions to a single number and it is followed by a dropout layer with probability $0.5$.
This architecture allows the input lattice to be of any size.
Finally, a dense layer with a single unit without activation is added to output the Casimir energy.
There is a $L_2$-regularization for all weights.
The gradient descent is performed with the Adam algorithm with the performance measured by the root-mean-square error, using a batch size of $32$ and early stopping (the maximum number of epochs is fixed to $200$, in practice it requires around $150$ before stopping).
Neural networks work best when all variables have similar scales: the output (energy) is normalized (subtraction of the mean and scaling to unit variance) and batch normalization is used between the intermediate layers.
Total, the neural network has circa $390$k parameters.
All these ingredients are standard~\cite{ref:review:1, Chollet:2017:DLP} and aim at making the learning faster and preventing overfitting and underfitting (improve generalization).
The code is written using Keras, an open-source neural-network library written in Python~\cite{ref:keras}.

\ed{We have randomly generated a few thousands of thin boundaries, which included closed non-self-intersecting lines and, separately, two quasi-parallel nonintersecting lines.} To sample different shapes and size scales, we generated the curves in a few independent runs so that the general distribution the dimensions of the curves does not correspond to a Gaussian. The number of samples for the different datasets is as follows:
\begin{itemize}
    \item $2d$, $(256, 256)$: $3000$ deformed circles, 
    $3000$ 
    lines;
    \item $3d$, $(255, 255)$: $2000$ deformed circles, $2000$ lines;
    \item $2d$, $(512, 512)$: $5000$ deformed circles.
\end{itemize}
The neural network is trained for each dataset separately.
In each case, the dataset is split in three sets: $80\%$ for training, $10\%$ for validation (to tune the parameters of the network) and $10\%$ for testing. In $3d$, the training takes circa 5 min for 800 samples (running on a GPU GeForce GTX 1080), while predicting takes circa 5 ms for one sample. For comparison, Monte Carlo takes 3.1 hours for a single sample on a GPU Tesla K40.

\begin{table}[ht]
\centering
\begin{tabular}{c|c|c|c|c|c}
    & \multicolumn{2}{c|}{$(255, 255)$} &
    \multicolumn{2}{c|}{$(256, 256)$}
    &      $(512, 512)$
    \\
	& $3d$ circles & $3d$ lines
	         & $2d$ circles & $2d$ lines
	         & $2d$ circles
	\\
	\hline
	samples & $200$ & $200$
	     & $300$ & $300$
         & $500$
	\\
	\hline
	mean
	     & $0.064$ & $0.0037$
	     & $0.048$ & $0.0025$
         & $0.084$
	\\
	min
	     & $0.000087$ & $0.000019$
	     & $0.00003$ & $0.000024$
         & $0.000047$
	\\
	75\%
	     & $0.069$ & $0.0051$
	     & $0.060$ & $0.0034$
         & $0.096$
	\\
	max
	     & $2.1$ & $0.016$
	     & $0.87$ & $0.015$
         & $1.1$
\end{tabular}
\caption{Relative errors for $3d$ (for the Casimir energy ${\cal E}_{\cal C}$) and $2d$ (for the mean action $\avr{S}$) compared to the MC result, evaluated for the deformed circles and the quasi-parallel lines. The line ``75\%'' gives the third quartile (75\% of the errors are below the value), ``min'' and ``max'' are the minimum and maximal errors.}
\label{tbl:errors}
\end{table}

\begin{figure}[!thb]
\begin{center}
\includegraphics[scale=0.55,clip=true]{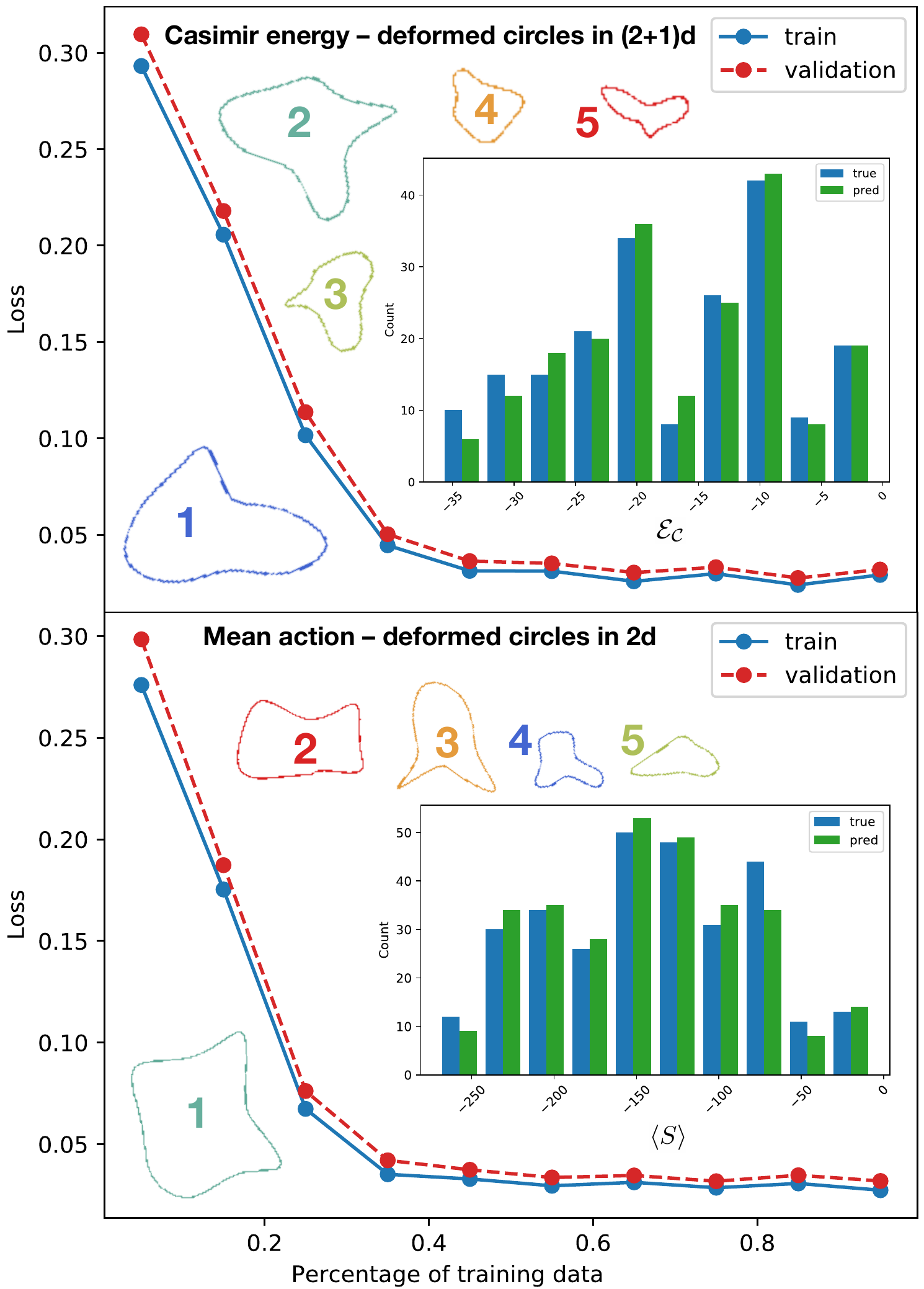}
\end{center}
\vskip -4mm 
\caption{Learning curves for (top) the Casimir energy ${\cal E}_{\cal S}$ in $3d$ scalar model and (bottom) the mean action $\avr{S}$ in $2d$ scalar model for the set of the deformed circles at $255^2$ and $256^3$ lattices, respectively. The inset histograms confront, statistically, the real vs.~predicted distributions of the Casimir energies ${\cal E}_{\cal S}$ and actions $\avr{S}$, for the $3d$ and $2d$ sets, respectively. Several examples of the deformed circles are shown as well (described in Table~\ref{tbl:examples};
See Supplemental Material at [URL will be inserted by publisher] for the examples of the deformed circles in a digital data format).}
\label{fig:results:closed}
\end{figure}

\begin{table}[ht]
\centering
\begin{tabular}{c|c|c|c|c|c|}
\cline{3-6}
\multicolumn{2}{c}{} &\multicolumn{2}{|c|}{MC} & \multicolumn{2}{|c|}{ML} \\
\cline{1-6}
 & $N$ & ${\cal O}$ & err${}_{\cal O}$ & ${\cal O}$ & err${}_{\cal O}$ \\
\hline
\parbox[t]{2mm}{\multirow{5}{*}{\rotatebox[origin=c]{90}{$3d$}}} 
& 1	&		-22.62	&	0.13	&	-22.60	&	0.014		\\
& 2 &	 	-20.34	&	0.12	&	-20.34	&	0.0018	\\
& 3	&		-12.22	&	0.09	&	-12.23	&	0.011	\\
& 4	&		-9.57	&	0.16	&	-9.57	&	0.0028	\\
& 5	&		-9.57	&	0.013	&	-9.56	&	0.011	\\
\hline
\parbox[t]{2mm}{\multirow{5}{*}{\rotatebox[origin=c]{90}{$2d$}}} 
& 1	&	    -227.44 &   0.43    &   -227.43		&	0.0068 \\
& 2	&		-165.06 &   0.42    &   -165.05		&	0.014	\\
& 3	&		-156.99 &   0.42    &   -156.98		&	0.018	\\
& 4	&	    -95.30  &   0.42    &   -95.38		&	0.019 \\
& 5	&	    -90.44  &   0.42    &   -90.45		&	0.088	\\
\hline
\end{tabular}
\caption{The Casimir energy (${\cal O} = {\cal E}_{\cal S}$) in the $3d$ model and the mean action (${\cal O} = \avr{S}$) in the $2d$ model obtained with the first-principle Monte-Carlo (MC) calculation and the Machine Learning (ML) techniques, together with appropriate absolute errors. The numbers $N$ label the deformed circles shown in Fig.~\ref{fig:results:closed}.}
\label{tbl:examples}
\end{table}

The inset histograms in Fig.~\ref{fig:results:closed} characterize the statistical features of the predictive power of the ML algorithm. The histograms show, in a statistical manner, the number of the deformed circles with given range of the Casimir energies in $3d$ (the mean action in $2d$) obtained with the help of the Monte-Carlo calculations (``true'') as compared to the predicted by the neural network (``pred.''). 
The errors are summarized in Table~\ref{tbl:errors}. For the majority of samples, the relative errors are small and the neural network reproduces well the MC result.
The largest errors are found for very small curves, as it can be expected (the image resolution is not sufficient for the neural network).
The learning curves represent the evolution of the root-mean-square error (loss) in terms of the number of samples used for training the neural network.
The validation data corresponds to all the data not used for training.
For large training sets, the flattening of both curves indicate that there is enough samples for training the network, the quasi-absence of gap between them shows that there is no overfitting, and the overall low values of the losses signals the absence of underfitting.
Together, this shows that the architecture of the network is well adapted to the task.

We also demonstrate the success of the method in Table~\ref{tbl:examples} for a set of particular examples, visualized and labeled in the insets of Fig.~\ref{fig:results:closed}. It is interesting to notice that in most cases the neural network gives the prediction very close to the mean actual value, which falls well within the errors both at Monte-Carlo and Machine-Learning sides. This fact, most probably, highlights a (cautionary) overestimation of the errors provided by the algorithms at the both sides.

We got very similar results for the learning curves, the statistical distribution and the magnitude of errors, for the set of quasi-parallel lines, with typical examples visualized in the right panel of Fig.~\ref{fig:nn-model} and relative errors presented in Table~\ref{tbl:errors}.

The predictive power of the ML algorithm depends on the size of the curve in dimensionless units (pixels). It seems that both for the coarser ($256^2$) and the finer-graded ($512^2$) lattices, there is a common scale $L$ (in pixels) below which the neural network does not work well; see Fig.~3 with the data shown in Table~III of Supplemental Material at [URL will be inserted by publisher] for the examples of these worst configurations. The existence of the lower pixel size needed to maintain the predictive power of the ML algorithm is naturally consistent with the expected property that for a fixed physical size of the curve, a finer discretization gives better results.

In our article, we demonstrated that the trained neural networks provide us with a quick and accurate tool for prediction of the zero-point energies of the physical --  though idealized, in our exploratory approach -- bodies. The methods are both versatile and universal as we successfully applied them in two physical setups (in $2d$ and $3d$) and for different types of boundaries (the deformed circles and the corrugated lines). The machine learning techniques may open the door to designing geometries with requested characteristics of the vacuum forces.

\begin{acknowledgments}
The authors are grateful to A.~N.~Chernodub (Grammarly) for useful comments. The numerical simulations were performed at the computing cluster Vostok-1 of Far Eastern Federal University. The work was supported by a grant of the Russian Foundation for Basic Research No. 18-02-40121 mega. H.E.\ was supported by a Carl Friedrich von Siemens Research Fellowship of the Alexander von Humboldt Foundation during most of this project. V.G. is partly supported by the Research Center for Nuclear Physics, Osaka University Collaboration Research network (COREnet).
\end{acknowledgments}


\appendix

\section{Supplemental Material}

For the overwhelming majority of the studied boundary geometries, the trained neural network gives very good results. However, there is a small subset of the configurations for which our method does not work. In the first part of this supplemental material, we provide some examples where the neural network gives the worst results in terms of relative errors. In Table~\ref{tbl:bad:examples} we compare the Casimir energy ${\cal E}_{\cal C}$ obtained with the help of the first-principles Monte Carlo (MC) simulations (configurations A, B and C for the $3d$ model) and the mean action $\avr{S}$ (configurations D, E and F for the $2d$ model) with the corresponding quantities predicted by the Machine Learning (ML) method. 

\begin{table}[ht]
\centering
\begin{tabular}{c|c|c|c|c|c|}
\cline{3-6}
\multicolumn{2}{c}{} &\multicolumn{2}{|c|}{MC} & \multicolumn{2}{|c|}{ML} \\
\cline{1-6}
 & $N$ & ${\cal O}$ & err${}_{\cal O}$ & ${\cal O}$ & err${}_{\cal O}$ \\
\hline
\parbox[t]{2mm}{\multirow{3}{*}{\rotatebox[origin=c]{90}{$3d$}}} 
& A   &	 	-0.82	&	0.12	&	-2.54	&	1.72	\\
& B   &		-1.63	&	0.10	&	-2.67	&	1.04	\\
& C   &		-1.48	&	0.09	&	-2.30	&	0.82   \\
\hline
\parbox[t]{2mm}{\multirow{3}{*}{\rotatebox[origin=c]{90}{$2d$}}} 
& D
&	    -6.71 &   0.42    &   -12.53		&	5.82  \\
& E   &		-14.44 &  0.42    &   -8.34		&	6.10	\\
& F   &		-41.15 &  0.43    &   -28.38		&	12.78   \\
\hline
\end{tabular}
\caption{Worst predictions (in terms of relative error) from the ML for the Casimir energy (${\cal O} = {\cal E}_{\cal S}$) in the $3d$ model and the mean action (${\cal O} = \avr{S}$) in the $2d$ model. The labels $N$ correspond to the deformed circles shown in Fig.~\ref{fig:bad:examples}.}
\label{tbl:bad:examples}
\end{table}

\begin{figure}[!h]
    \centering
    \includegraphics[scale=1.35]{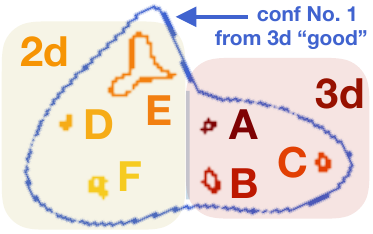}
    \caption{The examples A, B, $\dots$, F of the deformed circles for which the neural network makes the worst predictions (given in Table~\ref{tbl:bad:examples}). For comparison, we plot -- keeping the correct scale -- the configuration No. 1 from the good examples for the $3d$ model shown in the upper plot of Fig.~\ref{fig:results:closed}.}
\label{fig:bad:examples}
\end{figure}

The worst configurations A, $\dots$, F are visualized in Fig.~\ref{fig:bad:examples} in yellow/red colors. It turns out that these configurations correspond to relatively small loops, where the effects of the coarse-graining are large. We also visualize, for comparison, one of the good configurations which was already shown in the upper plot of Fig.~\ref{fig:results:closed}.

\begin{figure*}[thb]
    \centering
    \includegraphics[scale=0.825]{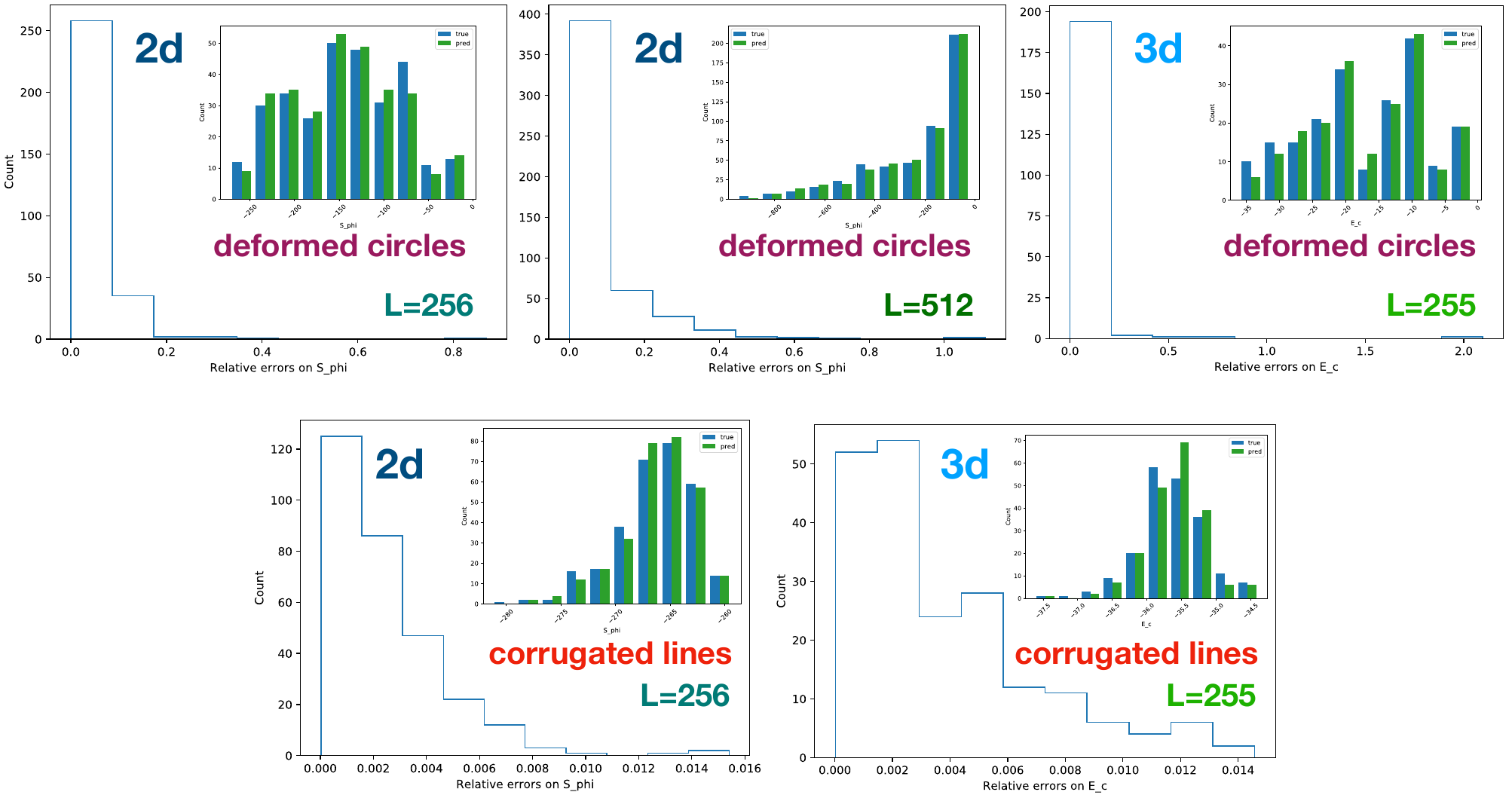}
\caption{Distribution of values (MC and ML) and errors.}
\label{fig:all-distrib}
\end{figure*}

For completeness, we also show in Fig.~\ref{fig:all-distrib} the distributions of the MC and ML values and of the ML relative errors (ML minus MC value divided by the MC value).
The bad examples described above are responsible for the tails in the error distributions for the deformed circles. Note, however, that there are very few such instances with high errors: for example, in the $2d$ case with $L = 256$, out of the total of $300$ samples, there is only one sample with error of circa $0.85$, one sample with the error about $0.4$, and two samples with errors approximately equal to $0.3$. On the other hand, these large relative errors appear for small curves which have a small value of total Casimir energy. As a consequence, the absolute error with respect to the MC result is small, which explains why the distributions of the MC and ML values agree very well (notice that the range of the values of the Casimir energies is quite large).

We also find that the relative errors are much smaller for the case of the corrugated lines. On the other hand, the range of Casimir energies for the lines is more restricted than the one for deformed circles, while the energy values are relatively large. Therefore, even small relative errors can be visible in the value distribution, as it is clearly seen for the $3d$ case of the corrugated lines in Fig.~\ref{fig:all-distrib}.

\end{document}